\documentclass[useAMS,usenatbib]{emulateapj}
\usepackage{epsfig}
\usepackage{times}

\newcommand{\ltsima} {$\; \buildrel < \over \sim \;$}
\newcommand{\gtsima} {$\; \buildrel > \over \sim \;$}
\newcommand{\lta} {\lower.5ex\hbox{\ltsima}}
\newcommand{\gta} {\lower.5ex\hbox{\gtsima}}

\begin{document}

\title{Identification and properties of the photospheric emission in GRB090902B}
\author{
F.~Ryde\altaffilmark{1,2}, 
M.~Axelsson\altaffilmark{3,2}, 
B.~B.~Zhang\altaffilmark{4},
S.~McGlynn\altaffilmark{1,2}, 
A.~Pe'er\altaffilmark{5},
C.~Lundman\altaffilmark{2,1}, 
S.~Larsson\altaffilmark{6,2}, 
M.~Battelino\altaffilmark{1,2}, 
B.~Zhang\altaffilmark{4},
E.~Bissaldi\altaffilmark{7},
J.~Bregeon\altaffilmark{8},
M.~S.~Briggs\altaffilmark{9}, 
J.~Chiang\altaffilmark{10}, 
F.~de~Palma\altaffilmark{11,12},
S.~Guiriec\altaffilmark{9},
J.~Larsson\altaffilmark{3,2},
F.~Longo\altaffilmark{13,14}, 
S.~McBreen\altaffilmark{7,15}, 
N.~Omodei\altaffilmark{12}, 
V.~Petrosian\altaffilmark{16},
R.~Preece\altaffilmark{9},
A.~J.~van~der~Horst \altaffilmark{17}
}
\altaffiltext{1}{Department of Physics, Royal Institute of Technology (KTH), AlbaNova, SE-106 91 Stockholm, Sweden}
\altaffiltext{2}{The Oskar Klein Centre for Cosmoparticle Physics, AlbaNova, SE-106 91 Stockholm, Sweden}
\altaffiltext{3}{Department of Astronomy, Stockholm University, SE-106 91 Stockholm, Sweden}
\altaffiltext{4}{Department of Physics and Astronomy, University of Nevada, 4505 Maryland Parkway, Las Vegas, Nevada 89154-4002, USA}
\altaffiltext{5}{Space Telescope Science Institute, 3700 San Martin Drive, Baltimore, MD 21218, USA ; Riccardo Giacconi fellow}
\altaffiltext{6}{Department of Physics, Stockholm University, AlbaNova, SE-106 91 Stockholm, Sweden}
\altaffiltext{7}{Max-Planck Institut f\"ur extraterrestrische Physik, 85748 Garching, Germany}
\altaffiltext{8}{Istituto Nazionale di Fisica Nucleare, Sezione di Pisa, I-56127 Pisa, Italy}
\altaffiltext{9}{Center for Space Plasma and Aeronomic Research (CSPAR), University of Alabama in Huntsville, Huntsville, AL 35899, USA}
\altaffiltext{10}{W. W. Hansen Experimental Physics Laboratory, Kavli Institute for Particle Astrophysics and Cosmology, Department of Physics and SLAC National Accelerator Laboratory, Stanford University, Stanford, CA 94305, USA}
\altaffiltext{11}{Dipartimento di Fisica ``M. Merlin" dell'Universit\`a e del Politecnico di Bari, I-70126 Bari, Italy}
\altaffiltext{12}{Istituto Nazionale di Fisica Nucleare, Sezione di Bari, 70126 Bari, Italy}
\altaffiltext{13}{Istituto Nazionale di Fisica Nucleare, Sezione di Trieste, I-34127 Trieste, Italy}
\altaffiltext{14}{Dipartimento di Fisica, Universit\`a di Trieste, I-34127 Trieste, Italy}
\altaffiltext{15}{University College Dublin, Belfield, Dublin 4, Ireland}
\altaffiltext{16}{Center for Space Science and Astrophysics, Stanford University, 382 Via Pueblo Mall, Stanford, CA 94305, USA}
\altaffiltext{17}{NASA Postdoctoral Program Fellow, NASA/Marshall Space Flight Center, 320 Sparkman Drive, Huntsville, AL 35805, USA}

\label{firstpage}

\begin{abstract}
The \textit{Fermi Gamma-ray Space Telescope} observed the bright and long GRB090902B, lying at a redshift of $z = 1.822$. Together the Large Area Telescope (LAT) and the Gamma-ray Burst Monitor (GBM) cover the spectral range from 8 keV to $>$300 GeV.  Here we show that the prompt burst spectrum is consistent with emission from the jet photosphere combined with nonthermal emission described by a single powerlaw  with photon index -1.9.
The photosphere gives rise to a strong quasi-blackbody spectrum which is somewhat broader than a single Planck function and has a characteristic temperature of $\sim 290$ keV.  We model the photospheric emission with a multicolor blackbody and its shape indicates that the photospheric radius increases at higher latitudes. We derive the averaged photospheric radius $R_{\rm ph}= (1.1 \pm 0.3) \times 10^{12} \, Y^{1/4} \mathrm{cm}$ and the bulk Lorentz factor of the flow, which is found to vary by a factor of two and has a maximal value of $\Gamma = 750 \, Y^{1/4}$. Here $Y$ is the ratio between the total fireball energy and the energy emitted in the gamma-rays. We find that during the first quarter of the prompt phase the photospheric emission dominates, which explains the delayed onset of the observed flux in the LAT compared to the GBM.  We interpret the broad band emission as synchrotron emission  at $R \sim 4 \times 10^{15}$ cm. Our analysis emphasize the  importance of having high temporal resolution when performing spectral analysis on GRBs, since there is strong spectral evolution.
\end{abstract}


\section{Introduction}

The mechanism giving rise to prompt emission in gamma-ray bursts (GRBs) has long been a puzzle. The emission typically peaks in the 100-1000 keV range and is modeled with the empirical Band function, which consists of two exponentially joined power-laws \citep{band}.  In general the narrow spectral ranges available during observations of GRBs have made it difficult to unambiguously determine the emission mechanism. This has been remedied by the launch of the {\it Fermi Gamma-ray Space Telescope} which regularly observes GRBs with its two instruments, the Large Area Telescope \citep[LAT, nominal energy range 20\,MeV--$>$300\,GeV;][]{lat} and the Gamma-ray Burst Monitor \citep[GBM, 8\,keV--40\,MeV;][]{gbm}, thus covering an unprecedented spectral range.

The bright, long GRB090902B was detected by {\it Fermi} and was one of the brightest observed to date by the LAT \citep{chi09}. Over 200 photons were detected at energies above 100\,MeV, including 39 photons with energy above 1\,GeV \citep{deP09}. The LAT also detected a photon with energy $33.4^{+2.7}_{-3.5}$\,GeV, the highest seen in any GRB so far. The burst lies at a redshift of $z=1.822$ \citep{cuc09}, which yields an isotropic energy of $E_{\rm iso}= (3.63 \pm 0.05) \times 10^{54}$ ergs, and the afterglow emission was seen at X-ray, optical, NIR and radio wavelengths. 

The prompt spectrum of GRB090902B over the  energy range 8 keV-33 GeV shows clear deviation from the generally expected Band function \citep{chi09}. The time-integrated spectrum is best fitted by the addition of a separate power-law component (photon index of $\sim -1.9$) to the Band function. The power-law is detected at both lower and higher energies around the Band component. During the first half of the prompt emission phase, the Band power-law index at energies below the spectral peak significantly violates the optically thin synchrotron limit; photon index of $\alpha = -2/3$ \citep{preece98,preece02}.

\cite{ryde05} suggested that GRB spectra are a superposition of two spectral components: photospheric blackbody emission and an accompanying nonthermal component. Over the limited energy range of 20-2000 keV (using data from BATSE on {\it CGRO}) studied, an adequate model consists of a single Planck function and a power-law. However, the nonthermal component, which is modelled by the power-law, is not well characterized, since the burst spectral shape is in general dominated by the thermal component (see further \citet{ryde08,rydepeer09}).  The \textit{Fermi} data now allow an improved characterization of the two components due to the increased energy range. In this paper, we investigate in more detail whether the observed two components in the broadband spectrum of GRB090902B could be attributed to such a photospheric model.

\section{Time-resolved spectral analysis}

\citet{chi09} analysed time-resolved spectra of GRB090902B with time bins of typical duration of  6\,s. 
In particular, they studied the behavior of the power-law component in the spectrum. In this work, we investigate the peaked component of the spectrum, i.e. the ``Band" component, at a higher time resolution in order to study the spectral evolution in greater detail. A consequence of this is that any model will be less constrained at LAT energies (cp. \cite{chi09}). Of particular interest is the first half of the prompt phase ($t = 0 - 13$\,s), during which the Band component is very hard and narrow, which challenges optically-thin emission mechanisms as its origin. The Band component has a low-energy power-law   $N_{\rm E} \propto E^{ \alpha} =  E^{0}$ and a high-energy power-law $N_{\rm E} \propto E^{ \beta} = E^{-4}$.   We therefore focus our study on epochs {\it a}, {\it b}, and {\it c} in \cite{chi09}, and only briefly comment on the later behavior. 

The time binning was chosen by requiring a signal-to-noise ratio of 40 in the most strongly illuminated GBM detector, NaI 1 (see Table 1). We also include data from NaI detector 0, BGO 0 and 1, and from the LAT in our fits.  The LAT ``transient" class data contain front and back events which are considered separately \citep{lat}.  The NaI data are fit from 8 keV to 1 MeV and the BGO from 250 keV to 40 MeV using the Time Tagged Event (TTE) data type. The LAT data are fit from 100 MeV.  An effective area correction factor of 0.9 is applied to the BGO with respect to the NaI detectors and LAT. The fits were performed with the spectral analysis software package RMFIT (version 3.0).

The first result we find is that the $\alpha$-value gets larger, i.e. the sub-peak spectrum gets harder when  narrower time intervals are used for the Band+powerlaw model. For the time intervals used in \cite{chi09} $\alpha \sim 0$; with the narrower time bins adopted here, we obtain an average value of $\alpha \sim+ 0.11$ (see Table 1). Several spectra have  $\alpha \sim + 0.3$, which are among the hardest GRB spectra ever measured \citep{ryde04,kaneko06}. We therefore attempt to 
interpret this hard and narrow spectral component as stemming from the photosphere. First we model it with  a single Planck function, which is used in addition to the powerlaw (BB+pl).  The average value of $C-stat/{\rm dof} =  634/599$ (reduced $C-stat = 1.058$). We note however that the distribution of the residuals between the model and data shows trends that are not expected from stochastic variations, and thus indicate that the peak is slightly broader than a single Planck function.

Indeed, from a theoretical point of view the photospheric emission is not expected to be a pure Planck function.
A broadening of the thermal component is expected due to contributions from different regions in space. 
Goodman (1986) discussed the emerging spectrum from an optically thick, fully ionized relativistic flow and showed that the emerging spectrum has a peak that is broader and a slope below the peak that is slightly shallower compared to the Planck function. Several effects needs to be taken into account. The observed  blackbody temperature depends on the latitude angle due to the angle dependence of the Doppler shift. Likewise, the optical depth is angle dependent, which results in the photospheric radius increasing with angle (Pe'er 2008). Angle dependent density profiles of the outflow will have a similar effect. Therefore, the photospheric component is better represented by a multicolor blackbody instead of a single Planck function. The broad energy range now available through the Fermi observations, allows us to model the spectrum of the photospheric emission in greater detail than previously since the accompanying nonthermal component can now better be constrained.  We thus fit the photospheric component with a phenomenological multicolor blackbody (mBB), which is  given by
\begin{equation}
F^{\rm mBB} (E, T_{\rm max}) = \int_{} ^{T_{\rm max}} \,\, \frac{dA(T)}{dT} \,\, \frac{ E^3}{exp[E/kT]-1}\,\, dT
\end{equation}
The spectrum thus consists of a superposition of  Planck functions in the temperature range $T = T_{\rm min}$ to $T_{\rm max}$; $T_{\rm max}$  is a free parameter ($T_{\rm min}<<T_{\rm max}$  and therefore cannot be determined). The spectrally integrated flux for each Planck function is given by $F(T) = A(T) \, T^4 \pi^4 / 15$, where $A(T)$ is the normalization. Phenomenologically, we introduce the index $q$, relating the flux and the temperature of the individual Planck  functions 
\begin{equation}
F(T) = F_{max} \left(  \frac{T}{T_{\rm max}} \right)^q,
\end{equation}
where $F_{max}  = F(T=T_{\rm max})$. 
This full thermal + nonthermal model has 5 free parameters; power-law index,  powerlaw normalization,  $T_{max}$, $A(T_{\rm max})$, and finally $q$.

\begin{figure*}
 \begin{center}
\epsfig{file=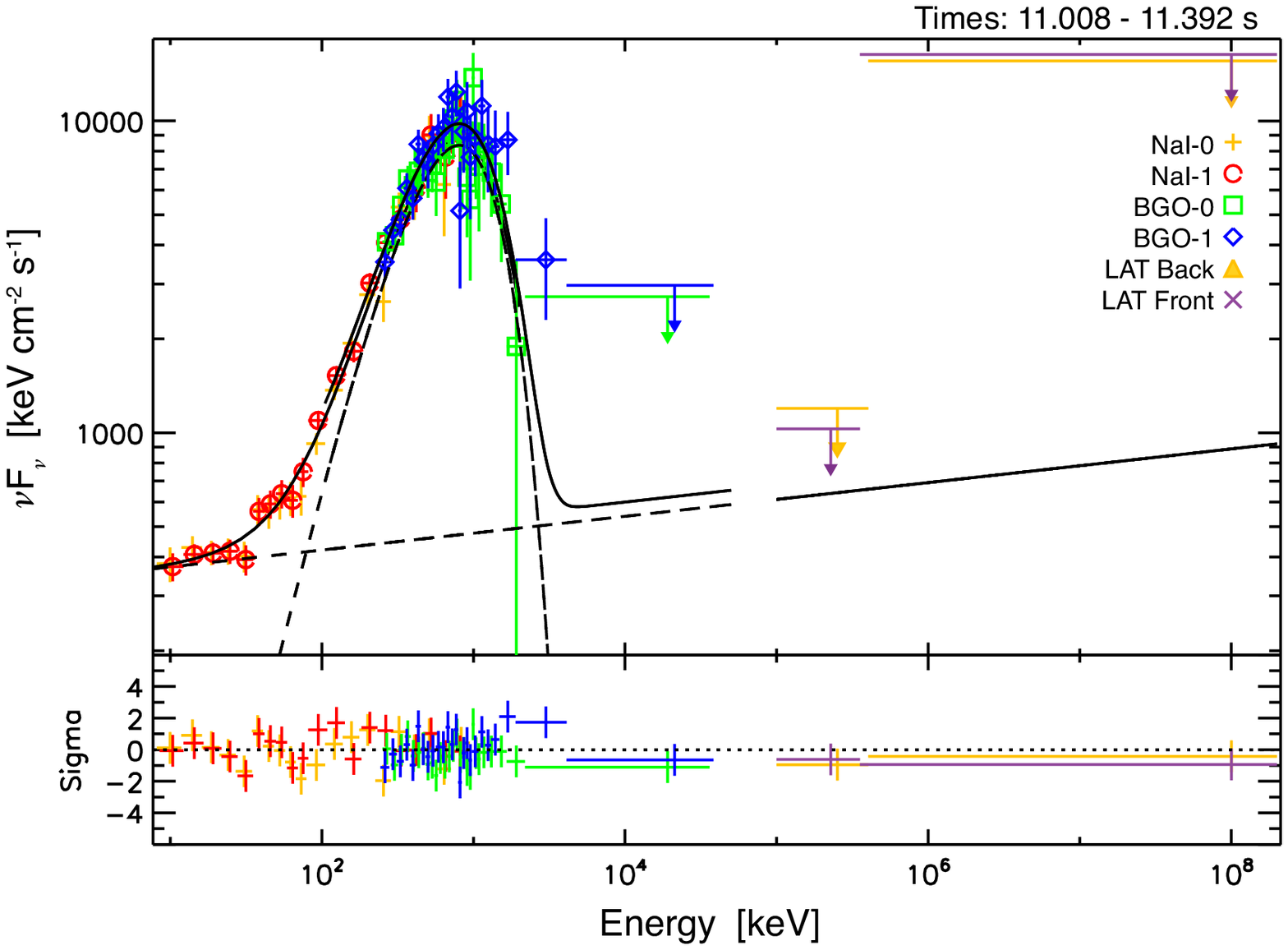,width=140mm}
 \end{center}
 \caption{Time-resolved $\nu F_\nu$ spectrum for the interval $t=11.008 - 11.392$ s over the GBM + LAT energy ranges fitted with a multicolor blackbody + power-law model.
 }
 \label{fig:3}

\end{figure*}

 Figure~\ref{fig:3} shows the time-resolved $E \, F_{\rm E} $ spectrum from one of the time intervals (11.008--11.392\,s after the GBM trigger). The mBB function captures the sharp spectral peak emerging above the nonthermal component, being a factor of $\sim 30$ above it. Such a strong thermal component is rarely seen in GRB spectra \citep{ryde04}. For this time-bin 
$kT_{\rm max} = 270.1^{+14.5}_{-13.5}$ keV, and ${\cal{R}} \equiv \left( F_{\rm bb} / \sigma T^4 \right)^{1/2} = (3.50 \pm 0.05) \times 10^{-19}$ \citep{peer07}. The power-law index of the nonthermal component is $-1.95^{+0.018}_{-0.013}$ and the $q$-parameter $2.00^{+0.14}_{-0.12}$. The fit has a $C-stat/{\rm dof} =  516/598= 0.86$. The quality of the fits {improve} compared to the BB+pl model. Over all time bins the averaged $C-stat/{\rm dof} = 0.944$, which is an improvement of C-stat with 69 for 598 {\rm dof}. Moreover, the residuals do not have any marked trends. 

\begin{figure*}
 \begin{center}
\epsfig{file=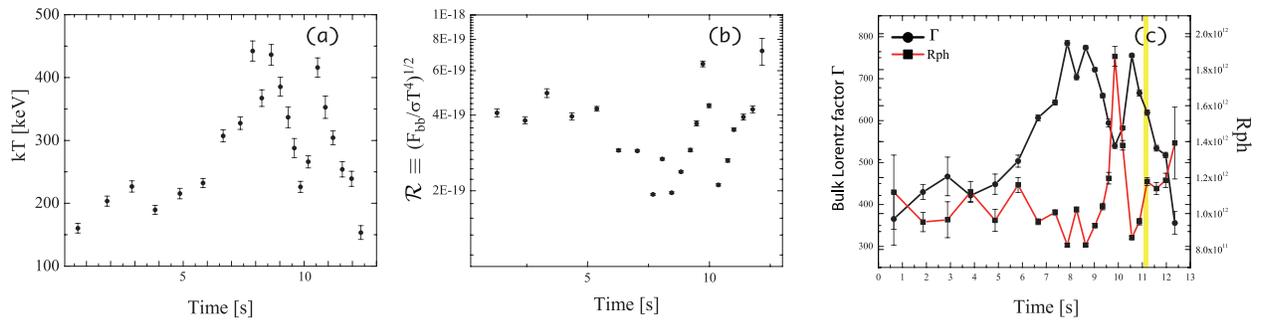,width=165mm}
 \end{center}
 \caption{Evolution of  (a) $kT$, (b)  ${\cal{R}} $,  
 (c) $\Gamma$ (black) and $R_{\rm ph}$ (red).  The yellow line indicates the time of the detection of the 11.21 GeV photon.}
 \label{fig:1b}
\end{figure*}

In Fig.~\ref{fig:1b} the evolution of the characteristic temperature is shown.  Due to the brightness of the burst and the relative strength of the thermal component, the temperature is determined with unprecedented accuracy. Since the burst consists of heavily overlapping pulses the characteristic cooling behavior over a single pulse \citep{rydepeer09} is not clearly detected. The middle panel in Fig.~\ref{fig:1b} shows the dimensionless parameter ${\cal{R}}$. For the studied interval in GRB090902B it is fairly constant, varying only by a factor of 2. Typically, ${\cal{R}}$ varies by an order of magnitude \citep{rydepeer09}. Finally, the $q$-parameter does not vary much during the analyzed interval and has an averaged value $q=1.9$ with a standard deviation of $0.4$.  

An alternative interpretation for the broadening of the thermal component is the effects of scattering and Comptonization, as worked out in \citet{peer05,peer06}. Moreover, heating of thermal electrons by Coulomb collisions with protons will give rise to a broadened  photospheric component  \citep{bel09}. We therefore also fit the photospheric component with a spectrum  taking these effects into account. This is approximated by a Band function with $\alpha=0.4$ and $\beta=-2.5$. The averaged $C-stat/{\rm dof} = 0.937$, is similar to the mBB model and the residuals do not have any marked trends.

The nonthermal component is adequately  fitted by a single power-law and  a curvature in the spectrum is not statistically required.
The photon index is constant,  $\sim -1.95$, with a standard deviation of 0.05. This value is largely independent of the width of the time bins as well as of model used for the thermal component \citep{chi09}.

The fraction of energy flux in the thermal emission relative to the total flux in the observed energy band is on average $70 \%$. During the first 6 s the average is $86 \%$ (the spectrum is dominated by the thermal component), after which it settles to an averaged value of $63 \%$. \cite{chi09} noted that there is a delayed onset of the LAT light curve compared to the GBM light curve. This can be naturally explained by the dominance of  the photospheric component during the first 6 s of the burst. To investigate this further we reduced the signal to noise ratio to 10, increasing the number of time intervals. The $q$-parameter was frozen to $1.9$ (the average value) and the data were fit with the photospheric component, with only two free parameters, the temperature and the normalization. The resulting, averaged reduced $C-stat$ has an  acceptable value of 1.02 for $t = 0 - 6$\,s.  Thereafter the reduced $C-stat$-value increases dramatically to $\sim 4 $ , indicating that an extra component is needed, i.e. the power-law component.  This is thus a natural explanation for the observed lack of photons in the LAT and in the NaI below 14 keV at the beginning of the pulse. We note that a similar behavior was observed by BATSE in GRB 970111 (Fig.13 in \citet{ryde04},\citet{ghirlanda04}).

Finally, we calculated the cross correlation function between the thermal and nonthermal light curves, which are found by integrating the mBB and power-law functions over the observed energy band for every time bin. We find a lag of $0.47\pm 0.27$ s (thermal emission leading) and a correlation maximum of $0.91 \pm 0.09$, indicating a strong correlation. 

Beyond the  time interval studied here, we showed in \citet{chi09} that the Band component becomes broader and softer. We find that the photospheric model (mBB) can adequately fit these spectra as well, however the value of  the $q$ parameter  is markedly lower, lying between 1 and 1.5: the properties of the photosphere changes. Further theoretical study of the interpretation and reason for this is underway. 

We thus conclude that by resolving the light curve on a sub-second timescale one can identify a photospheric component in the spectrum, combined with a single power-law component. The photospheric component can be modelled by a hard and narrow Band function. A physical interpretation is given by a multicolor blackbody, which has one parameter less to be fit.  Using wider time bins, as done in \citep{chi09}, will increase the SNR but the variations in temperature and flux will broaden the spectrum,
weakening the thermal signature.  Our results thus emphasize the  importance of having the highest possible temporal resolution when performing spectral analysis on GRBs.
\subsection{Properties of the photosphere}

The identification of the emission from the photosphere allows us to determine physical properties of the relativistic outflow, such as the bulk Lorentz factor $\Gamma$, the photospheric radius $R_{\rm ph}$  and the initial size of the flow $R_0$,  as we showed in \citet{peer07}.
The photospheric radius is given by $R_{\rm ph} = L \sigma_{\rm T}/8\pi \Gamma ^3 m_{\rm p} c ^3$, where $L$ is the total fireball luminosity, and $\sigma_{\rm T}$ and $m_{\rm p}$ are the Thomson cross section and proton mass respectively. The values of $R_{\rm ph}$ and $\Gamma$ can be determined by combining this expression with the measurement of the  dimensionless parameter ${\cal{R}}$, which is related to the effective transverse size of the photosphere
\begin{equation}
{\cal{R}}= \xi \, \frac{(1+z)^2}{d_{\rm L}} \, \frac{R_{\rm ph}}{\Gamma},
\end{equation}
where  $\xi$ is a geometrical factor of order unity, and  $d_{\rm L}$ is the luminosity distance.  Furthermore, we can estimate the 
radius above which the relativistic acceleration begins, $R_{\rm 0}$. This is done by combining  classical, nondissipative fireball dynamics with the fact that the total fireball luminosity $L = 4 \pi d_{\rm L} ^2 Y F_{\rm obs}$, where $Y$ is the ratio between the total fireball energy and the energy emitted in the gamma-rays, where $F_{\rm obs} = F_{\rm bb} + F_{\mbox{\scriptsize non-th}}$.

The discussion in \citet{peer07} assumed a single Planck spectrum. However the discussion can be generalized to slightly distorted photospheric spectra, as in the ones discussed above. For instance, the $q$-value we find here ($q\sim 2$) indicates that at a certain observer time the photospheric radius increases with latitude angle, so for every time bin we therefore {estimate an average} photospheric radius. The values found below are similar to the ones found by  approximating the spectrum with a single Planck function and applying the theory directly.

All the parameters evolve with time. In particular we find that the bulk Lorentz factor $\Gamma$ starts off at a value of close to  $\Gamma \sim 550$ for the first 6 s. Thereafter it rises sharply to a maximal value of $\Gamma \sim 750$ \citep[cf.][]{chi09}.
The  time-averaged value  (with standard deviation)  for  the bulk Lorentz factor is $\Gamma = 580 \pm 130 \left( \xi Y  \right)^{1/4}$,
while the time-averaged photospheric radius is found to be $R_{\rm ph}= (1.1 \pm 0.3) \times 10^{12} \mathrm{cm}\, \xi^{-3/4} Y^{1/4}$, see the right panel in Figure \ref{fig:1b}. The photospheric radius is remarkably stable compared to the variations in temperature and flux. Furthermore, since $\Gamma$ is most strongly dependent on the temperature, the evolution of these quantities track each other. We also estimate the radius at which the jet is launched to $ R_{\rm 0} = (1.0 \pm 0.5) \times 10^{9} \, \xi^{-4} Y^{-3/2} \mathrm{cm} $, and consequently the saturation radius $ R_{\rm s} \equiv \Gamma R_0 = (5.2 \pm 1.7) \times 10^{11}\,\,\, \xi^{-15/4} Y^{-5/4} \mathrm{cm} $. Here we  assume  nondissipative fireball dynamics. However, we note that the dynamics depends on the magnetization  \citep{giannios09} and on the presence of subphotospheric heating, such as tangential collimation shocks \citep{rm05,lazz09}, and  Coulomb heating \citep{bel09}.

\section{Discussion}

The presence of a strong photospheric component in GRB spectra has been discussed by several authors \citep[e.g. ][]{good86,MRRZ,rm05}\footnote{see also e.g. \citet{pac86,mes00,LU,DM,DS,R,GS,peer06,ruffini}}. Here we have identified a strong photospheric component in GRB090902B, which is modelled by a multicolor blackbody.   It dominates over  the nonthermal component, being nearly 100\% at early times. 

Varying energy injection at the central engine, or variations caused by interaction with the progenitor material  can cause the observed variability in the light curve. The flow is advected through the photosphere, where the thermal emission escapes. A fraction of the kinetic energy stored in the flow is later dissipated and emitted as nonthermal emission, e.g. synchrotron radiation. It is therefore expected that both the thermal and nonthermal emission reflect the properties and original energy injection at the central engine and thus are correlated with each other. This is consistent with the high correlation observed,  with a time lag of approximately 0.5 s. Such a lag corresponds to a shock radius $R_{\rm sh} - R_{\rm ph} = 2 \, c \Gamma^2 t_{\rm lag} (1+z)^{-1} \sim 4 \times 10^{15} (\Gamma/580)^2   \mathrm{cm} \sim R_{\rm sh}$ at which the optical depth $\tau \sim (R_{\rm sh}/R_{\rm ph})^{-2} \sim 10^{-7}$ ensuring an optically thin-emission site. 

\cite{chi09} estimated a lower limit on the bulk flow Lorentz factor $\Gamma_{\rm min}  \sim 1000$  by combining the energy of the 11.16 GeV photon, observed at 11 s after the GBM trigger, with the variability timescale  in the LAT data which was determined to be $t_{\rm v} \sim 0.1$ s: The gamma-ray opacity for pair production should be less than unity in order to allow the photons to escape without attenuation from the flow. This  Lorentz factor  is slightly higher than the value we determined above for the flow advected through the photosphere. However, the estimations of $\Gamma$ based on opacity arguments are sensitive to uncertainties in determining the variability time  \citep{zhangpeer09}.  Indeed, assuming that the variability timescale in the nonthermal component is set by the angular timescale at the  shock radius, we have $t_{\rm v} = t_{\rm ang}^{sh} =   (1+z)^{} \,\,R_{\rm sh}/ (2 \, c \Gamma^2)= 0.5$ s.  This value is consistent with the  fact that there is not much variability strength on timescales shorter than 1 s, as measured by the power density spectrum of the LAT data.   A variability timescale of  0.5 s yields  $\Gamma_{\rm min} \sim 750$, which is close the value we estimate. Finally, we note that the observed variability timescale in the GBM data should  reflect the angular timescale of the photosphere. The determined value $t_v^{\rm GBM} \sim 50$ ms \citep{chi09} is larger than $(1+z)^{} \,\, R_{\rm ph}/ (2 \, c \Gamma^2)\sim  2 \times 10^{-4}$ s, which is to be expected as shown by Pe'er (2008). 
    
As shown above, the photospheric emission allowed us to estimate $R_0$ assuming nondissipative acceleration: $ R_{\rm 0} \sim10^{9}\, \xi^{-4} Y^{-3/2} \mathrm{cm} $.   This is larger than the physical size  around a solar mass black hole, typically assumed to be at the centre of the GRB engine, with a Schwarzschild radius of $R_{\rm sch} = M/c^2 \sim  10^7 \mathrm{cm}$ \citep{pac86}. 
This might be an indication of that the fireball dynamics includes significant dissipation during the acceleration phase and/or subphotospheric heating, which would lower the estimated value of  $ R_{\rm 0} $. Alternatively, the larger size we find might correspond to the radius of the stellar core, at which the jet is launched. \cite{thom07} argue that internal dissipation within the star prevents the Lorentz factor from significantly rising until the jet escapes the core of the progenitor star, such as a Wolf-Rayet star. However,  \cite{lazz09}) showed that the dissipative shocks in the jet continue far beyond the stellar radius, thereby  influencing the jet dynamics.

Some bursts have no strong evidence of a narrow thermal component. For cases like GRB080916C which has a standard Band spectrum covering 6-7 orders of magnitude, the lack of a thermal component can be taken as an argument for a Poynting flux dominated flow \citep{zhangpeer09}. For other cases that show a standard Band spectrum in a narrower energy range, like during the second half of the prompt phase in 090902b, it is possible that a modified photosphere spectrum with subphotospheric heating can account for the data \citep{peer07,bel09,lazz2009}. Broadband observations of more GRB prompt emission spectra are desirable to more definitely diagnose the composition of GRB jets. 

\section*{Acknowledgments}

The $Fermi$ LAT Collaboration acknowledges support  for development and operation of the LAT, and scientific data analysis from NASA and DOE (USA) , CEA/Irfu, IN2P3/CNRS, and CNES (France), ASI, INFN, and INAF (Italy), MEXT, KEK, and JAXA (Japan), and the K.A.W. Foundation, VR and SNSB (Sweden). We thank Drs. Beloborodov and Lazzati for useful discussions. AP is supported by the Riccardo Giacconi fellowship at STScI and AJvdH by the NASA/MSFC Postdoc-program.

\newpage

\begin{table*}
\begin{minipage}{146mm}
\caption{Spectral fits.}
\begin{tabular}{l|lcccccrr}
\hline\hline
 Time & Model & PL Index & $\alpha$ & $q$/$\beta$ & kT$_{max}$/$E_{\rm pk}$  &  C-stat & Red. C-stat  \\ 
(s) & & & & & (keV) & /DoF & \\
\hline
& & & & & & &\\
0.00--1.28 & mBB+PL & $-2^{+1}_{-unc}$ & & $1.8^{+0}_{-0.2}$ & $161^{+9}_{-13}$  & 500/598 & 0.84 \\ 
 & Band+PL & $-2.15^{+0.17}_{-0.64}$ & $0.03^{+0.14}_{-0.15}$ & $-2.9^{+0.5}_{-0.3}$ & $403^{+24}_{-27}$ & 497/597 & 0.83 \\
1.28--2.43 & mBB+PL & $-2.00^{+0.20}_{-0.10}$ & & $1.8^{+unc}_{-0.1}$ & $204^{+11}_{-12}$  & 555/598 & 0.93 \\
 & Band+PL & $-2.10^{+0.25}_{-unc}$ & $-0.03^{+0.12}_{-0.09}$ & $-3.5^{+0.5}_{-unc}$ & $545^{+27}_{-26}$ & 556/597 & 0.93 \\
2.43--3.33 & mBB+PL & $-1.80^{+21.4}_{-0.10}$ & & $1.79^{+unc}_{-0.09}$ & $218^{+unc}_{-31}$  & 573/598 & 0.96 \\
 & Band+PL & $-1.84^{+0.09}_{-0.12}$ & $-0.06^{+0.11}_{-0.10}$ & $-3.0^{+0.2}_{-0.3}$ & $563^{+32}_{-31}$ & 561/597 & 0.94 \\
3.33--4.35 & mBB+PL & $-1.85^{+0.08}_{-0.07}$ & & $2.0^{+unc}_{-0.2}$ & $190^{+10}_{-11}$  & 535/598 & 0.89 \\
 & Band+PL& $-1.87^{+0.07}_{-0.09}$ & $0.14^{+0.11}_{-0.12}$ & $-3.2^{+0.3}_{-0.6}$ & $502^{+25}_{-23}$ & 531/597 & 0.89 \\
4.35--5.38 & mBB+PL & $-1.90^{+0.10}_{-0.10}$ & & $1.7^{+unc}_{-0.1}$ & $216^{+unc}_{-12}$  & 723/598 & 1.21 \\
 & Band+PL & $-1.94^{+0.11}_{-0.12}$ & $-0.12^{+0.10}_{-unc}$ & $-9.9^{+5.8}_{-unc}$ & $572^{+50}_{-26}$ & 722/597 & 1.21 \\
5.38--6.27 & mBB+PL & $-1.90^{+0.10}_{-0.10}$ & & $2.1^{+unc}_{-0.1}$& $233^{+11}_{-10}$  & 668/598 & 1.12 \\
 & Band+PL & $-1.94^{+0.10}_{-0.09}$ & $0.21^{+unc}_{-0.10}$ & $-4.9^{+1.2}_{-unc}$ & $649^{+25}_{-26}$ & 669/597 & 1.12 \\
6.27--7.04 & mBB+PL & $-1.90^{+0.05}_{-0.05}$ & & $2.2^{+0.1}_{-0.1}$ & $307^{+15}_{-16}$  & 570/598 & 0.95 \\
 & Band+PL & $-1.91^{+0.06}_{-0.05}$ & $0.28^{+0.10}_{-0.11}$ & $-3.4^{+0.5}_{-0.3}$ & $824^{+36}_{-40}$ & 564/597 & 0.94 \\
7.04--7.68 & mBB+PL & $-1.85^{+0.02}_{-0.02}$ & & $2.2^{+0.1}_{-0.1}$ & $317^{+15}_{-16}$ & 564/598 & 0.94 \\
 & Band+PL & $-2.08^{+0.12}_{-0.28}$ & $0.26^{+0.12}_{-0.17}$ & $-2.8^{+0.2}_{-0.2}$ & $818^{+42}_{-40}$ & 524/597 & 0.88 \\
7.68--8.06 & mBB+PL & $-1.92^{+0.04}_{-0.03}$ & & $1.9^{+0.1}_{-0.1}$ & $442^{+23}_{-25}$  & 502/598 & 0.84 \\
 & Band+PL & $-1.95^{+0.04}_{-0.05}$ & $0.04^{+0.10}_{-0.10}$ & $-3.6^{+0.3}_{-0.5}$ & $1179^{+58}_{-57}$ & 493/597 & 0.86 \\
8.06--8.45 & mBB+PL & $-2.02^{+0.05}_{-0.04}$ & & $1.9^{+0.1}_{-0.1}$ & $367^{+18}_{-19}$  & 486/598 & 0.81 \\
 & Band+PL & $-2.02^{+0.04}_{-0.05}$ & $0.07^{+0.11}_{-0.10}$ & $-4.0^{f}$ & $976^{+42}_{-539}$ & 493/598 & 0.82 \\
8.45--8.83 & mBB+PL & $-1.92^{+0.03}_{-0.03}$ & & $1.8^{+0.1}_{-0.1}$ & $436^{+23}_{-25}$  & 560/598 & 0.94 \\
 & Band+PL & $-1.93^{+0.04}_{-0.04}$ & $-0.03^{+0.10}_{-0.09}$ & $-4.7^{+1.0}_{-14.5}$ & $1174^{+56}_{-58}$ & 557/597 & 0.93 \\
8.83--9.22 & mBB+PL & $-1.93^{+0.03}_{-0.03}$ & & $2.0^{+0.2}_{-0.1}$ & $358^{+21}_{-20}$ & 600/598 & 1.00 \\
 & Band+PL & $-1.94^{+0.03}_{-0.04}$ & $0.12^{+0.12}_{-0.11}$ & $-4.4^{+0.8}_{-1.9}$ & $1058^{+50}_{-53}$ & 598/597 & 1.00 \\
9.22--9.47 & mBB+PL & $-1.96^{+0.03}_{-0.04}$ & & $1.9^{+0.2}_{-0.2}$ & $313^{+26}_{-24}$ & 511/598 & 0.85 \\
 & Band+PL & $-2.00^{+0.04}_{-0.05}$ & $0.13^{+0.17}_{-0.16}$ & $-3.1^{+0.3}_{-0.4}$ & $870^{+66}_{-59}$ & 499/597 & 0.84 \\
9.47--9.73 & mBB+PL& $-2.08^{+0.06}_{-0.07}$ & & $1.8^{+0.2}_{-0.2}$ & $267^{+19}_{-18}$  & 484/598 & 0.81 \\
 & Band+PL & $-2.08^{+0.06}_{-0.07}$ & $0.03^{+0.16}_{-0.15}$ & $-5.4^{+1.6}_{-unc}$ & $774^{+42}_{-39}$ & 486/597 & 0.81 \\
9.73--9.98 & mBB+PL & $-2.04^{+0.05}_{-0.05}$ & & $2.0^{+0.2}_{-0.2}$ & $210^{+12}_{-12}$ & 511/598 & 0.86 \\
 & Band+PL & $-2.05^{+0.05}_{-0.06}$ & $0.18^{+0.15}_{-0.14}$ & $-4^{f}$ & $613^{+28}_{-26}$ & 511/598 & 0.85 \\
9.98--10.37 & mBB+PL & $-2.01^{+0.04}_{-0.04}$ & & $2.2^{+0.2}_{-0.2}$ & $248^{+14}_{-13}$ & 532/598 & 0.89 \\
 & Band+PL & $-2.02^{+0.04}_{-0.05}$ & $0.26^{+0.13}_{-0.12}$ & $-4^{f}$ & $735^{+31}_{-29}$ & 530/598 & 0.89 \\
10.37--10.75 & mBB+PL & $-1.94^{+0.03}_{-0.02}$ & & $2.0^{+0.1}_{-0.1}$ & $387^{+23}_{-21}$ & 588/598 & 0.98 \\
 & Band+PL & $-1.96^{+0.04}_{-0.04}$ & $0.08^{+0.11}_{-0.10}$ & $-4.0^{+0.5}_{-1.3}$ & $1123^{+60}_{-61}$ & 582/597 & 0.97 \\
10.75--11.01 & mBB+PL & $-1.98^{+0.05}_{-0.06}$ & & $1.7^{+0.1}_{-0.1}$ & $327^{+23}_{-22}$ & 522/598 & 0.87 \\
 & Band+PL & $-1.99^{+0.05}_{-0.06}$ & $-0.14^{+0.13}_{-0.12}$ & $-4.7^{+1.2}_{-unc}$ & $931^{+55}_{-51}$ & 521/597 & 0.87 \\
11.01--11.39 & mBB+PL & $-1.95^{+0.02}_{-0.01}$ & & $2.0^{+0.1}_{-0.1}$ & $270^{+15}_{-14}$ & 516/598 & 0.86 \\
 & Band+PL & $-2.07^{+2.02}_{-0.05}$ & $-0.03^{+0.11}_{-0.09}$ & $-4.6^{+2.5}_{-unc}$ & $820^{+32}_{-37}$ & 673/597 & 1.13 \\
11.39--11.78 & mBB+PL & $-2.06^{+0.06}_{-0.09}$ & & $1.9^{+0.2}_{-0.2}$ & $236^{+16}_{-15}$ & 541/598 & 0.91 \\
 & Band+PL & $-2.08^{+0.07}_{-0.09}$ & $0.04^{+0.14}_{-0.13}$ & $-3.8^{+0.6}_{-3.6}$ & $690^{+38}_{-37}$ & 540/597 & 0.90 \\
11.78--12.16 & mBB+PL & $-2.02^{+1.38}_{-0.04}$ & & $2.2^{+0.3}_{-0.2}$ & $222^{+14}_{-16}$ & 669/598 & 1.12 \\
 & Band+PL & $-1.99^{+0.07}_{-0.05}$ & $0.30^{+0.20}_{-0.18}$ & $-2.9^{+0.19}_{-0.12}$ & $643^{+40}_{-38}$ & 508/597 & 0.85 \\
12.16--12.54 & mBB+PL & $-1.89^{+0.13}_{-0.05}$ & & $0.9^{+0.1}_{-unc}$ & $141^{+16}_{-15}$ & 537/598 & 0.90 \\
 & Band+PL & $-2.46^{+0.43}_{-1.57}$ & $-0.79^{+0.27}_{-0.21}$ & $-2.4^{+0.1}_{-0.2}$ & $263^{+38}_{-30}$ & 522/597 & 0.87 \\
 \hline
\end{tabular}

$f=$fix, $unc=$unconstrained
\end{minipage}
\end{table*}

\label{lastpage}

\end{document}